Smartphone-Based Test and Predictive Models for Rapid, Non-Invasive, and Point-of-Care Monitoring of Ocular and Cardiovascular Complications Related to Diabetes

Kasyap Chakravadhanula[1]

1 - BASIS Scottsdale, 10400 N 128th St, Scottsdale, AZ 85259

Corresponding Author: Kasyap Chakravadhanula, (480)-825-6678

**Abstract**

Diabetes is a massive global problem, with growth especially rapid in developing regions, which can lead to several damaging complications. Among the most impactful of these are diabetic retinopathy, the leading cause of blindness among working class adults, and cardiovascular disease, the leading cause of death worldwide. However, diagnosis is often too late to prevent irreversible damage caused by these linked conditions. This study describes the development of an integrated test, automated and not requiring laboratory blood analysis, for screening of these conditions. First, a random forest model was developed by retrospectively analyzing the influence of various risk factors (obtained quickly and non-invasively) on cardiovascular risk. Next, a deep-learning model was developed for prediction of diabetic retinopathy from retinal fundus images by a modified and re-trained InceptionV3 image classification model. The input was simplified by automatically segmenting the blood vessels in the retinal image. The technique of transfer learning enables the model to capitalize on existing infrastructure on the target device, meaning more versatile deployment, especially helpful in low-resource settings. The models were integrated into a smartphone-based device, combined with an inexpensive 3D-printed retinal imaging attachment. Accuracy scores, as well as the receiver operating characteristic curve, the learning curve, and other gauges, were promising. This test is much cheaper and faster, enabling continuous monitoring for two damaging complications of diabetes. It has the potential to replace the manual methods of diagnosing both diabetic retinopathy and cardiovascular risk, which are time consuming and costly processes only done by medical professionals away from the point of care, and to prevent irreversible blindness and heart-related complications through faster, cheaper, and safer monitoring of diabetic complications. As well, tracking of cardiovascular and ocular complications of diabetes can enable improved detection of other diabetic complications, leading to earlier and more efficient treatment on a global scale.

*Keywords: Diabetic Retinopathy Screening, Smartphone Ophthalmology, Cardiovascular Risk, Point-of-Care Screening, Machine Learning, Computer Vision*

**1. Introduction**

Approximately four hundred and twenty million people worldwide have been diagnosed with diabetes mellitus, and this is predicted only to increase. Of those with diabetes, approximately one-third are expected to be diagnosed with diabetic retinopathy (DR), a chronic eye disease that can progress to irreversible vision loss, and the leading cause of blindness among working-class individuals (Facts About Diabetic Eye Disease, 2015). Early detection, which is vital for effective prognosis, relies on skilled readers and is both labor and time-intensive, which limits who can be helped. Skilled readers perform prognosis by analysis of swelling in the retina that threatens vision, of evidence of poor retinal blood vessel circulation, or of abnormal vessels or tissue in the retina [4]. Moreover, the manual nature of DR screening methods creates inconsistency among readers [4].

Xie et al. [15] found that patients with proliferative diabetic retinopathy have an increased risk of incident cardiovascular disease (which can manifest in abnormalities in ocular blood vessels), meaning that these patients must be followed up with more closely to prevent cardiovascular disease. Due to this connection, simultaneous monitoring has several advantages to improve early detection and risk prediction of incident cardiovascular disease.

Cardiovascular disease is the leading cause of death in the world, for both men and women. In the U.S., 1 in every 4 deaths is due to cardiovascular disease. Although cardiovascular disease is a leading cause of deaths globally, many do not act on risk factors and warning signs although about half (47%) of Americans have at least one significant risk factor [7].

For these extremely prevalent problems, current methods (detailed below) of risk detection/prediction are costly, slow, inaccessible, and often inconsistent. Due to the emergence of more accessible machine learning, machine learning and deep learning are being applied to various medical problems as well. Thus far application of these methods to automate prediction of diabetic retinopathy and cardiovascular risk, while accurate, have been resource-intensive and computationally costly, creating an inability to apply these methods on devices in low-resource settings. As well, these tests have been separated (see "Discussion" section for further comparison). This study aimed to apply various machine learning/deep learning algorithms to more cheaply, quickly, accessibly, and consistently monitor diabetic retinopathy and cardiovascular disease in a comprehensive test.

A recent publication by EyePACS [3] revealed the use of retinal fundus images in the prediction of diabetic retinopathy. Research has shown that many other aspects of health may be predicted via the analysis of retinal fundus images (age, sex, systolic BP, and smoker % among others), which are all risk factors of cardiovascular disease.

Although there exist fairly accurate automated tests to predict groups at risk of cardiovascular disease (for example Pooled Cohort and Framingham), these tests are time-consuming, resource-intensive, and invasive. This is due to the fact that several risk factors in the analysis are determined in a blood test, which often take much time to return results, are costly, and are often not available in low-resource areas [11, 12]. For this reason, the current method of assessing cardiovascular risk is not suitable for effective continuous monitoring on a large scale. The gap for a more accessible and easily performable test is the one this study desired to fill, by creating a smartphone-based, easily administered test through the analysis of both risk factors (via user input) and of retinal fundus images.

Current methods of diagnosing diabetic retinopathy (D.R.) require a retinal fundus image which must be taken in a properly equipped facility by a trained professional using a device that can cost ~$5,000. The image must then be processed by a professional reader, which is very time-consuming (up to 7 weeks). The current method is slow, inconvenient, expensive, and inconsistent (due to manual reading), and is thus ineffective for early detection and response before damage is irreversible [4]. In developing nations, where diabetic retinopathy is most damaging, the current solution cannot be implemented effectively.

Currently, there exist solutions to automate prediction of diabetic retinopathy from retinal fundus images via deep learning [4, 14], however current approaches either falter in accuracy or are too computationally resource intensive to be implemented on most mobile devices (see Discussion section for further explanation).

This study sought to construct an inexpensive, convenient, and consistent device for rapid retinal imaging and diagnosis, which can also be implemented effectively in low-resource settings. Two specific goals of this study are high accuracy in the prediction of cardiovascular risk without incorporating features that require laboratory blood analysis to determine, and a computationally light infrastructure leveraging existing models on the smartphone to enable both rapid and complex analysis on a mobile device. These are especially relevant in low resource settings (implementation in these settings with regard to adoption and technology availability is discussed further in "Conclusions").

Because cardiovascular disease accounts for 1 in 3 deaths in the US [7], and diabetic retinopathy is a leading cause of vision impairment and blindness worldwide [8], this study has enormous potential for societal impact.

**2. Materials and Methods**

**2.1 Cardiovascular Risk Factors Model**

Eleven input features (which could be taken non-invasively, quickly, and easily) and one output feature (the risk classification) were chosen from the University of California Irvine (UCI) Heart Disease Dataset [9]). The raw data came in the form of a spaced list of numbers, with random indents. VBA allowed this data to be structured in a .csv file. The final step was to use Pandas to convert the .csv to a python NumPy array, which could be used in the model. The "SciKit-learn" library was used to build models for this task. In order to choose a model structure, the pre-optimization testing accuracies (using the default parameters from Sci-Kit Learn) of several model types versatile on relatively small datasets were measured. Accuracy was measured by randomly isolating 20% of the dataset, allowing the model to train on the remaining 80%, and measuring the classification accuracy of the model on the 20% left for testing samples which the model had not yet seen. The 3 top performing structures (Random Forest, Classical Naive Bayes, K-Nearest Neighbors) are compared in Figure 1. The accuracy of the Random Forest model in this preliminary test was the highest, so the Random Forest Model was chosen for this task. The other models (Naïve Bayes, K-NN) were the 2nd and 3rd highest performing models and are depicted only for the sake of comparison; they have no further role in this study.

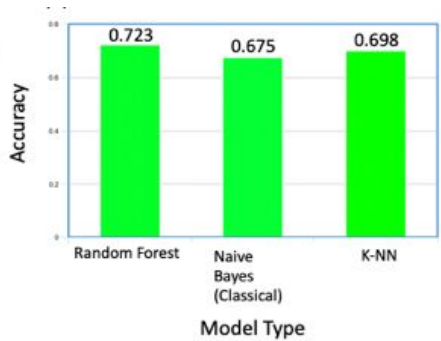

**Figure 1.** The accuracies of Random Forest, Classical Naïve Bayes, and K-Nearest Neighbor compared. Using a "voting" system, random forest classifiers can quickly learn on data and accurately validate predictions, leading to a higher efficiency and end performance.

The dataset was split randomly into training and testing sets (20% for testing, the rest is left for training). The optimized model was built in python with the "SciKit-learn" library and used n=100 estimators (the tree "nodes"), arrived at through iterative testing of each model in a range of 600 values of n. To measure the optimized testing accuracy of the model in a low-bias manner, the technique of k-fold cross validation was used. This technique creates k unique sets of training and testing data (where k is an integer, meaning the number of folds). To evaluate this model, k=10 folds were created, meaning 10 untrained copies of the model were trained and tested on unique, randomized splits of the original dataset. Over these 10 folds, the model structure reached an overall test accuracy of ~82%, which was higher than the goal accuracy and is comparable to other similar tests (discussed further in Discussion section).

**2.2 Diabetic Retinopathy Model**

Transfer Learning is the retraining of the final layers of a pre-trained and tested complex deep learning network, which results in a comparable accuracy, efficiency, and loss, to building the model from scratch. It requires much less training data, computational power, model size, and training time.

The pretrained model chosen was Inception v3, which showed higher preliminary accuracies than model structures as VGG and MobileNet. Inception v3 is a model trained to recognize images of 1000 different objects (classes). In transfer learning, the final "output" layer was retrained to return diagnosis of diabetic retinopathy via Google's Tensorflow.

The dataset chosen was the EYEPacs retinal fundus image dataset [13], a widely-used dataset published as part of an open competition which contains about 45,000 retinal fundus images and their corresponding grades.

Due to the complexity of the eye, our initial approaches to feed the raw images into a transfer deep learning network yielded low accuracies of below 60%, indicating need for revision.

Medically, experts diagnose diabetic retinopathy and other conditions in the eye through the observation of blood vessels. So, the model can safely treat all but the blood vessels as "noise" for the identification of diabetic retinopathy. To simplify the input data and remove this "noise," automatic vessel segmentation was used.

To validate the improvement that automatic vessel segmentation could have, the transfer learning model was tested on the DRIVE database [2], which has images which were manually vessel segmented and their corresponding diagnosis of diabetic retinopathy (graded by medical professionals). The model reported accuracy scores of almost 97%, showing the potential of this method.

However, the problem which still remained was the automatic "conversion" of retinal fundus images to segmented ones. Due to the complexity of the problem and the plethora of research already done on the many existing datasets, an open source contrast model (like that described in [1]) was transfer learned for this task on parts of the STARE database [6].

This model was tested on the datasets DRIVE [2], STARE [6], and HRF [5], which combined have hundreds of image pairs. The model's predictions closely mirrored the segmented images.

There is a visible bias for the lower values in the dataset because the higher levels are just much rarer in the world. However, if one class is overrepresented, especially in deep learning, the model can overfit and predict in accordance with the dataset rather than the image content.

A solution is increasing the number of data points at the higher levels. This is achieved by conducting data augmentation. Simple data augmentation is the inversion of images or the addition of noise to create new data and also create model "tolerance" for distortion. This was performed for the 3/4 class images through a simple python script. Figure 2 shows the effect of data augmentation on a fundus image.

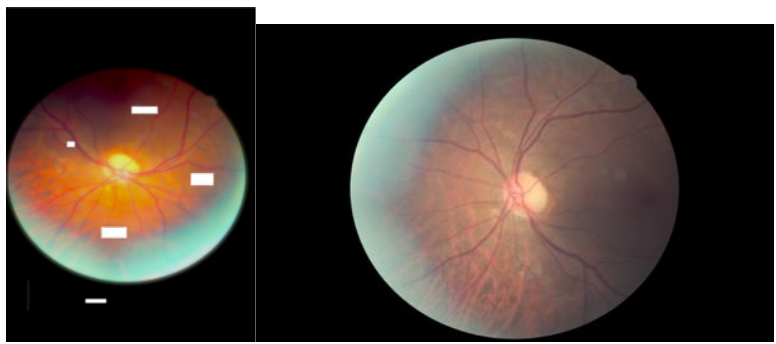

**Figure 2.** Left – Original Retinal Fundus Image, Right – Augmented Retinal Fundus Image with orientation changed, distortion added, alpha and relative exposure changed, and spots removed (Retinal Image from EYEPacs Dataset in [13], described above).

The final model architecture is shown in Figure 3.

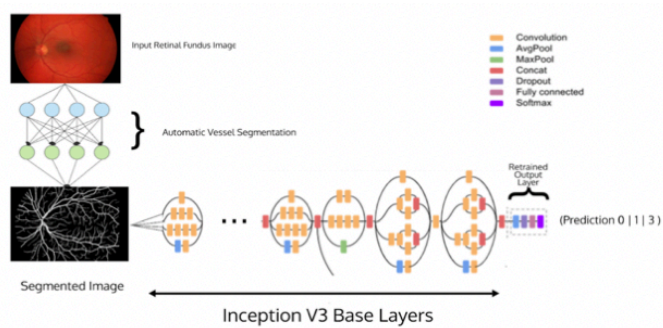

**Figure 3.** Final Model Architecture, with Automatic Vessel Segmentation and Inceptionv3 transfer learning. The model, when tested on the EYEPACS dataset using the 10-fold cross validation method described earlier, recorded a testing accuracy of **~81%**, which is highly respectable (see Discussion section for further comparison).

**2.3 Smartphone Implementation and Prototype**

The App and User Interface were coded/designed in the IDE XCode, in the Swift language.

Since the models were written in languages and with libraries not iOS native, the models were converted to CoreML, the iOS native construction for machine learning, using another library called coremltools. After this conversion, the app could make full use of the models. The next step was designing a professional and clean User Interface (UI) and User Experience (UX), which was achieved using the versatile XCode IDE.

While each individual model had been tested, the comprehensive test had yet to be evaluated.

The model was tested in multiple trials, on 50 random data collections from the UCI Heart Disease Dataset (each data collection consists of 12 values – 11 risk factors and 1 risk value). The UCI Heart Disease Dataset does not include retinal fundus images, so an image from the EyePACS dataset was inputted based on the D.R. classification of severity in the UCI collection.

Overall, the test had an accuracy of 80% over 50 collections. When the classification was made binary for each model, as it will likely be in a real-world implementation, the accuracy was 96%.

Because retinal fundus images are typically taken on an expensive and immobile machine (see Figure 4), a more mobile and cheaper version needed to be developed to fit the smartphone application.

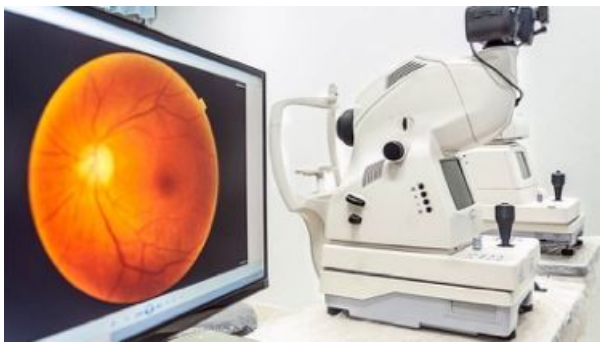

**Figure 4.** Traditional Retinal Imaging Machines, costing around $5,000.

11For this purpose, a 3D-printed smartphone attachment was designed and built. This attachment fits to the smartphone camera and is adjustable. As well, it is lightweight, compact, user-friendly (with some basic instruction), versatile, and inexpensive (about 100 times lower cost). The attachment can use the phone's native flash for lighting or any other coaxial light source.

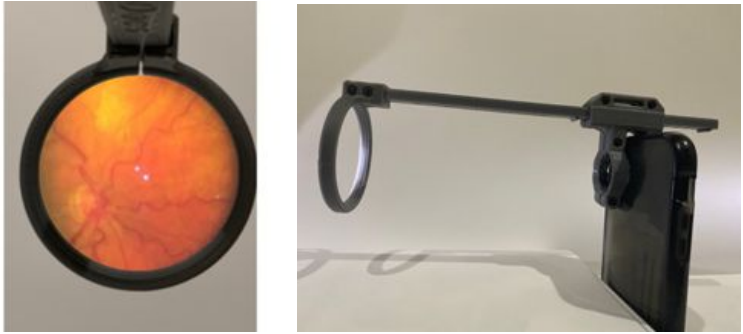

**Figure 5.** The left image shows the device camera view from a retinal fundus image (device tested on and image taken from real retina, background replaced with white). The right image shows the 3D-printed device fitted to the smartphone, which is the device configuration (lens was removed).

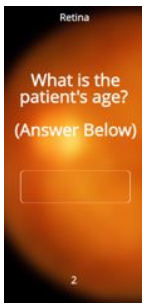

**Figure 6.** The app takes some basic risk factors as input for the models.

**2.4 Evaluation**

Although an important way to gauge model performance is simply model accuracy, there are several other, more descriptive ways to do so which ensure the model performs on multiple fronts.

**Receiver Operating Characteristic (ROC) curve**

ROC curves show the ability of a model to distinguish between classes. An ROC curve is created by varying the threshold between classifications and measuring the "True" or correct positive portion versus the "false" or incorrect positive portion. A good ROC curve means that the "overlap" between the classes is small (or the model is more "sure" in its predictions), especially important in medical computing problems.

The distance between the ROC curve and the centerline signifies the amount of distinctness the model assigns to each class (a high distance is more "sureness" and less class overlap).

In an ROC curve, the AUC ROC (Area under the ROC curve) is the best measure of model performance. An excellent model has an AUC ROC near to 1 (a horizontal line at TPR = 1), meaning it has a very high measure of separability. A poor model has an AUC ROC near to 0 (a horizontal line at TPR = 0), meaning it has the worst measure of separability. When the AUC ROC = 0.5, the model has no distinctive capability.

**Learning Curve (Test and Training Accuracies against Proportion of Data Set)**

A learning curve is a visualization of how training and test accuracies vary when the size of the dataset changes, as the model gains more experience. The learning curve is useful in that it shows how adding more data may or may not benefit the model, telling the researcher if pursuing more data will in fact be useful. It also is an indicator of overfitting and underfitting. The learning curves for both models are below (Figure 8).

The most important performance measure discerned from the learning curve is fit. There are two possible problems related to fit: overfit and underfit.

Overfit occurs when a model's predictive ability corresponds too closely to a particular dataset and may therefore fail to fit additional data or generalize to future observations reliably. Overfit is analytically when accuracy substantially drops with new data, such as in the testing set. If the curve for training (red) is not convergent with the testing line (green), the model has overfit, because the testing accuracy will never reach the training accuracy even with a large enough dataset.

In addition, if the curves greatly reduce in concavity with higher and higher values, the model is struggling to generalize to more data and noise, which is underfit.

**Feature Importance**

Feature importance is an important visualization for models with a possible medical application, largely because both physicians and patients need to trust the model, and also because feature importance analysis can reveal new insight on the condition or biological system. This graph can be used to inform patients and physicians about why the model made a specific prediction and to inform patients how to improve their risk classification, with the most important risk factors being age, rest heart rate, and classification of severity of diabetic retinopathy.

**3. Results**

**3.1 Receiver Operating Characteristic (ROC) curve**

Figure 7 shows the Receiver Operating Characteristic (ROC) curves for each model.

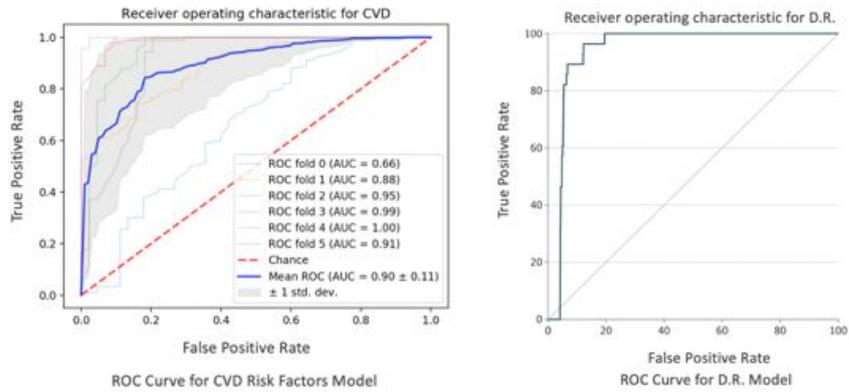

**Figure 7.** ROC curves for each model. The independent variable is the threshold value, and the dependent variable is the relationship between TPR and FPR.

The Mean AUC ROC (in 5-fold cross validation) of the cardiovascular risk factors model was 0.90 with 95% C.I. +/- 0.11, and the Mean AUC ROC (also in 5-fold cross validation) of the diabetic retinopathy model was 0.982 with 95% C.I. +/- 0.002.

Note: The data was dichotomized for this curve (because of the nature of the ROC measuring two-class distinction), so these curves demonstrate the models' distinction purely between at risk or not at risk.

**3.2 Learning Curve**

Figure 8 shows the Learning Curve for each model.

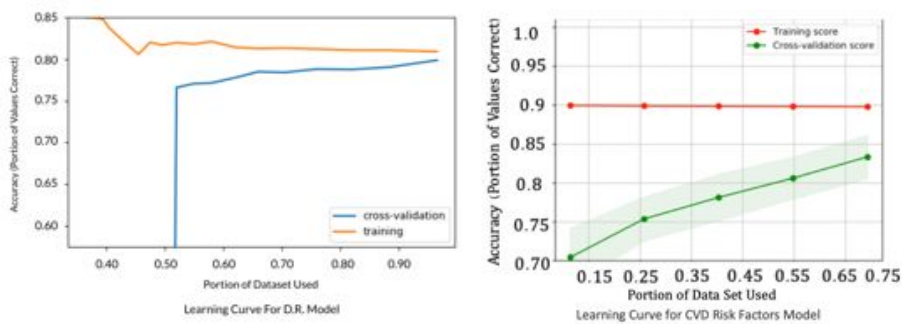

**Figure 8.** Learning curves for each model, with training indicated by the red curve (orange for D.R. model) and testing or cross-validation by the green (blue for the D.R. model). The independent variable is the data set portion, and the dependent variable is accuracy.

For the cardiovascular risk factors model, the shared point of convergence with more data shows the growth potential with more data. The cap is about 90%, which is extremely well-performing. With the dataset at hand, the model has not overfit but would benefit from more data. The model also has not underfit, which is positive and shows the potential of the model to generalize even better with more data.

For the diabetic retinopathy model on the EyePACS dataset, the model has not overfit and the point of convergence is roughly the same, which shows the model is almost at peak generalization. The model has seemed to underfit slightly, likely due to the complexity of the problem and the number of distinct images, along with the fact that the model was transfer learned.

**3.3 Model Visualization**

The feature importance graph is in Figure 9 for the cardiovascular risk factors model, with risk factors associated with their relative importance to the model prediction.

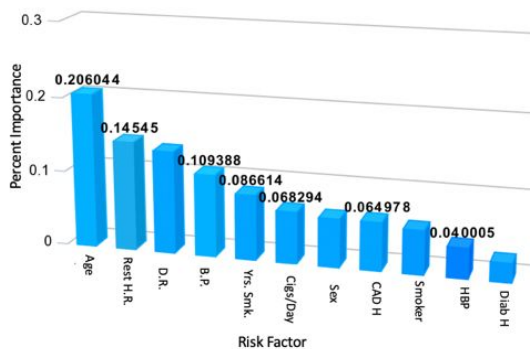

**Figure 9.** Relative importances of the chosen cardiovascular risk factors, represented graphically. Results obtained post-training of the random forest model.

4. Discussion

As stated in the introduction section, there exist deep learning approaches for automatic classification of diabetic retinopathy in retinal fundus images. The average AUC ROC of the transfer-learned model presented in this study is 0.982 (95% C.I., 0.980-0.984). In a study done by Google researchers classifying the same EYEPacs dataset in [13] using deep learning, an AUC ROC of 0.991 (95% C.I., 0.988-0.993) was determined [14]. In a separate study using deep learning to classify the same dataset, an AUC ROC of 0.951 (95% C.I., 0.947-0.956) was measured [17]. The model in [14] had a slightly higher AUC ROC, while that measured in [17] was significantly lower. Both models, upon trial, were not able to be implemented on the smartphone device used in this study along with several other recent smartphone models without cloud processing support due to resource needs of the complexities of the trained structures. Along with not being able to run analysis, trackage of energy and storage usage revealed significant feasibility issues on mobile devices upon trial. The model presented in this study freezes most of the trained layers in the original Inception v3 network, reducing resource-intensive complexity and allowing efficient prediction on a variety of smartphones in a rapid manner (with an average processing time of 4 seconds). In contrast to the more resource intensive models presented in [14, 17], many current smartphone-enabled approaches to automated diabetic retinopathy detection via deep learning methods have reported comparatively low accuracies. For example, in [10], a set of patients with known type 2 diabetes underwent imaging for mobile analysis. Out of these, the presence of diabetic retinopathy was only correctly detected automatically in 68.6% of patients. In comparison, the model presented in this study conducts this binary classification with 96% accuracy, and a classification of severity with above 80% accuracy. In a screening setting (with a binary classification), the device presented in this study is more effective than current methods.

Currently, automated cardiovascular risk prediction requires risk factors which are obtained via a blood test. A goal of this study was to remove this requirement to make risk assessment more accessible and rapid. In [11], a machine learning model was constructed to predict and identify people at risk of cardiovascular disease using 473 variables, including risk factors related to blood assays, diet and nutrition, health and medical history, family history, sociodemographic factors, psychosocial factors, physical activity, lifestyle, and physical measures. Predictive performance was assessed using the area under the receiver operating characteristic curve (AUC-ROC), which was reported to be 0.774 with a 95% Confidence Interval of 0.768-0.780. This improves on previous methods such as the Framingham score [12], however it requires many risk factors which are difficult to obtain in most regions globally. This prevents effectiveness on a large scale, especially in low-resource settings, meaning many people do not understand their risk. The model presented in this study replaces many of these risk factors with a consideration of the retinal fundus image of the patient and reports a mean AUC-ROC of 0.9 (95% CI ± 0.11). This is a statistically significant improvement. However, it is important to consider that the model presented in [11] was evaluated on data from many more participants. It is likely that the predictive performance of the model presented in this study will decrease with such a large representation, and future testing on a larger sample size must be conducted to more properly compare the models and evaluate the ability of the model presented in this study to generalize to other sub-populations. The UCI dataset is relatively small (about 1000 samples), however due to the screening nature of this application, the high cross validation accuracy, and the high AUC ROC reported above, the model remains effective in improving understanding about cardiovascular risk in an accessible, non-invasive, and quick way.

Xie et al. [15] found that patients with proliferative DR have an increased risk of incident cardiovascular disease, meaning that these patients must be followed up more closely to prevent cardiovascular disease. Current methods for the monitoring cardiovascular risk and diabetic retinopathy are separated, while the device presented in this study integrates monitoring of the cardiovascular and ocular complications of diabetes, leading to improved early intervention and treatment to prevent irreversible blindness and cardiovascular disease which can lead to death. Also, due to its improved convenience, cost, speed, and non-invasiveness compared to current methods, this device has the potential to make a much broader impact, especially in developing regions and on a global scale.

## 5. Conclusions

The new test demonstrates performance in several metrics which compare to or surpass those of cutting-edge technologies. Current devices and tests for diabetic retinopathy require expensive machines and trained doctors, whereas this device is much cheaper (the retinal imaging attachment costs about $30 to print), also indicating large economic profitability which would accelerate the application of this device. As well, the test is much faster (a few seconds for this test versus 2-7 weeks for current methods), meaning that early intervention is improved and that much more frequent monitoring is possible, which is lacking currently.

Current tests for cardiovascular risk (even automated ones) require a blood test which is expensive, invasive (which can be unsafe, especially for elderly patients or for those in regions where contamination rates are high), and time-consuming. All of these limit effective early treatment and make continuous monitoring impossible. This device instantaneously and cheaply predicts cardiovascular risk without the need for an inconvenient and invasive blood test. As well, current tests for these conditions are independent and infrequent, while in reality they should be tracked together with high frequency, due to their fundamental connectivity.

**Implementation in Developing Regions**

In communities around the world, these improvements mean that more people can easily understand and act on heart problems, saving lives across the world. Especially in developing nations, understanding complications in the heart is vital to a healthy life, and this technology shows potential to massively improve the ability of people in low-resource communities across the globe to detect and treat problems in their heart.

These combined, along with increased convenience and ease of use, increase the ability of this device to improve treatment for diabetic retinopathy and cardiovascular disease in developing nations. Additionally, the device facilitates continuous monitoring and remote analysis with the cloud, meaning that doctors can improve early intervention and administration of treatment even if they are not in the same region as the patient. By tracking and monitoring the progression of diabetes as it manifests in the eye and in the cardiovascular complications, we can improve early detection and intervention to detect and mitigate the complications of diabetes, even beyond the two focused on in this work. On a broader scale, this has special potential to improve the quality of medical care in developing nations with a shortage of doctors.

Smartphone adoption is currently relatively low in many developing regions and can vary wildly. In these settings, aid organizations can utilize this application on standard smartphones in order to screen large populations. For individual monitoring, donation to local doctors and hospitals who prescribe the devices to patients is the most effective strategy. Because 3D-printing is not yet widely accessible in developing nations, at first the attachments would be cheaply printed by aid organizations and their partners for deployment in low-resource communities. We have built collaborations with such aid organizations as OneWorld Health and Microsoft 4Afrika to explore deployment of our devices in low-resource regions across Africa.

As such, the device would, at least initially, be most effective in an assistive and screening role, conveying information and prediction to doctors to enhance treatment and ability to respond early before damage is irreversible.

**Limitations**

Despite promising results, the device has limitations which must be addressed.

The field of view (FOV) of the smartphone retinal imaging attachment is limited compared to professional retinal cameras, resulting in lower descriptiveness and lower resulting accuracy. However, this can be mitigated by introducing lower-quality images into the training set, which was done in this study but should be furthered. As well, the models need to be further tested on these lower FOV retinal images to determine the extent of the lowering of accuracy.

Due to this limitation, currently the device is most effective as an assistive aid to be paired with a test by a professional machine, if it is necessary. However, the device is still effective for continuous monitoring and early detection to signal the need for further testing if the condition is severe.

Furthermore, although there was a high number of images used, they may not represent external complications which affect diagnosis, lowering model accuracy. This is especially true for the cardiovascular risk factors models, as the number of samples (~1000) was much lower than that for the diabetic retinopathy model. More data is needed for both models to improve the device in this aspect, to ensure that the models do not falter when introduced with an unknown complication.

Finally, some features in the dataset used were self-reported (such as years smoker, approx. cigarettes per day, and family history of diabetes), and may be biased or incorrect. Although this limitation much be acknowledged, it is inherent to most health-related datasets and is compensated for by other predictive factors.

Nonetheless, the device shows promising results and a real potential to greatly improve monitoring and early detection of diabetic retinopathy and related cardiovascular risk, improving treatment and allowing intervention before damage is irreversible in the form of complete blindness or even death. Additionally, the convenience, rapidity, non-invasiveness, and low cost of this point-of-care device allows widespread implementation in developing nations to improve the accessibility to and quality of medical care.

**6. Future Directions**

**6.1 Model Improvement**

Due to the constraint of the lack of stronger computational capabilities for this study, model size and computational power needed were important factors in determining the correct approach, which naturally came at the cost of some accuracy. With extended computational resources and more time to train, this constraint would be eased, and higher model performance could be achieved. This will ensure maximum accuracy and efficiency before the models are tested in the real world. A higher number of datapoints will yield a more generalizable and therefore helpful model, so a goal of this study moving forward is the acquisition, validation, and use of more data.

**6.2 Preliminary Trial Feedback**

Another point of future work after the above steps is the gathering of data from preliminary assistive clinical trials in a closed setting where the patients are aware of the testing. This would measure the performance in a real-world system where real-world factors and noise come into play, and the effect on model performance as well as overall performance could be quantitatively as well as qualitatively measured. Potential problems with the test, the app, either model, the 3D-printed imaging attachment, or the deployment would be exposed and corrected before more widespread deployment. Specifically, the ability of the 3D-printed retinal imaging attachment to image a variety of distinct eyes would be evaluated. Currently, testing of this device, while present, has not been systematic to different conditions and sub-populations. This would lead to further improvement of the application based on real-world feedback. As well, both physicians and patients can slowly begin to gain trust in the system in a no-cost environment, also advancing the use of machine learning and point-of-care technologies in medicine in general.

**6.3 Integration with Acetone Sensor**

The device presented in this work intends to improve detection and monitoring of diabetic retinopathy and cardiovascular risk. Both of these conditions are related to diabetes; diabetic retinopathy is a direct complication of diabetes and diabetes has clearly been shown to be a prime risk factor for cardiovascular disease by several published studies, including [16]. In an effort to further improve early detection of diabetic retinopathy and cardiovascular disease, a novel saliva acetone sensor and accompanying smartphone app have been developed separate from this study (depicted in Figure 10).

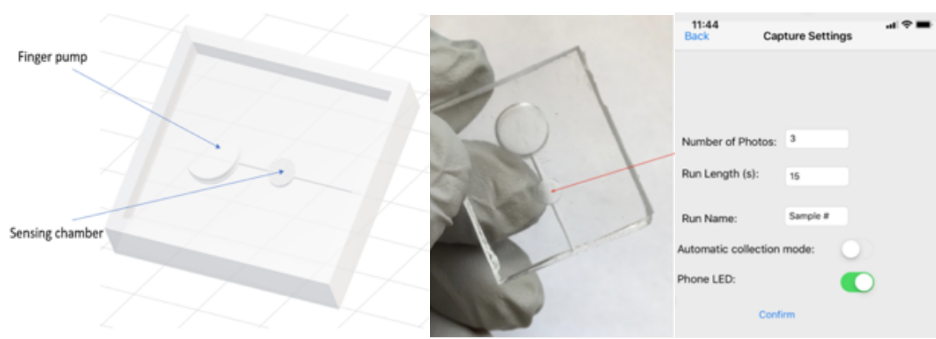

Figure 10. Left – Computer rendering of acetone sensor design, with Finger Pump and Sensing Chamber marked. Middle – Acetone sensor printed in Polydimethylsiloxane substrate. Right – Smartphone application to automatically take and analyze images.

23Acetone is well-known to be directly connected to the blood glucose levels of a patient (the traditional marker for diabetes). The sensor detects acetone through observation of a chemical reaction involving acetone (Figure 11) which produces a color change in the sensing chamber. By reading the red absorbance of the sensing chamber, acetone concentration can be determined (Figure 12). Spikes in the saliva concentration of acetone are good indicators of diabetic progression and thus of diabetic complications.

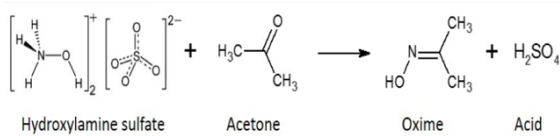

Figure 11. Hydroxylamine sulfate in the substrate reacts with acetone in the saliva sample, lowering the pH and triggering a color change in the pH indicator Thymol Blue.

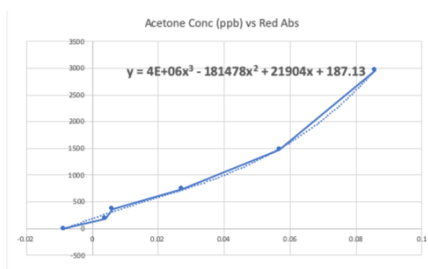

Figure 12. Predictive curve and equation to convert red absorbance of sensing chamber to sample acetone concentration.

The implementation of this sensor, integrated with the work in this study, may improve the ability of doctors to predict diabetic complications and respond early. While initial results are promising, the device and application are currently undergoing more comprehensive evaluation to measure sensitivity, accuracy, and potential issues.

**Acknowledgement of Role of Funding Source**

No external funding source to declare.

## 7. Summary

The goal of this work was to create an integrated test, automated and not requiring laboratory blood analysis, for screening for diabetic retinopathy and cardiovascular risk. First, a random forest model was developed by retrospectively analyzing the influence of various risk factors (obtained quickly and non- invasively) on cardiovascular risk. Next, a deep-learning model was developed for prediction of diabetic retinopathy from retinal fundus images by transfer learning the InceptionV3 model and pre-processing the images via automatic vessel segmentation. The models were integrated into a smartphone-based device, combined with an inexpensive 3D-printed retinal imaging attachment. Accuracy scores, as well as the receiver operating characteristic curve, the learning curve, and other gauges, were promising. This test is much cheaper and faster, enabling continuous monitoring for diabetes and its complications. It has the potential to replace the manual methods of diagnosing both diabetic retinopathy and cardiovascular risk, which are time consuming and costly processes only done by medical professionals away from the point of care, and to prevent irreversible blindness and heart-related complications through faster, cheaper, and safer monitoring of diabetes. Tracking of cardiovascular and ocular complications of diabetes can also enable improved detection of other diabetic complications, leading to earlier and more efficient treatment on a global scale.

## 8. References

1. G. Azzopardi, N. Strisciuglio, M. Vento, N. Petkov, Trainable cosfire filters for vessel delineation with application to retinal images, Medical image analysis 19 (2015) 46–57

2. J.J. Staal, M.D. Abramoff, M. Niemeijer, M.A. Viergever, B. van Ginneken, "Ridge based vessel segmentation in color images of the retina", *IEEE Transactions on Medical Imaging*, 2004, vol. 23, pp. 501-509.

3. Cuadros J, Bresnick G. EyePACS: An Adaptable Telemedicine System for Diabetic Retinopathy Screening. Journal of diabetes science and technology (Online). 2009;3(3):509-516

4. Lam C, Yi D, Guo M, Lindsey T. Automated Detection of Diabetic Retinopathy using Deep Learning. *AMIA Jt Summits Transl Sci Proc*. 2018;2017:147-155. Published 2018 May 18

5. Jan Odstrcilik, Jiri Jan, Radim Kolar, and Jiri Gazarek. Improvement of vessel segmentation by matched filtering in colour retinal images. In IFMBE Proceedings of World Congress on Medical Physics and Biomedical Engineering, pages 327 - 330, 2009.

6. A. Hoover, V. Kouznetsova and M. Goldbaum, "Locating Blood Vessels in Retinal Images by Piece-wise Threhsold Probing of a Matched Filter Response", *IEEE Transactions on Medical Imaging*, vol. 19 no. 3, pp. 203-210, March 2000.

7. Heart Disease Facts & Statistics. (n.d.). Retrieved from https://www.cdc.gov/heartdisease/facts.html

8. Facts About Diabetic Eye Disease. (2015, September 01). Retrieved from https://nei.nih.gov/health/diabetic/retinopathy

9. Janosi, Andreas & Steinbrunn, William & Pfisterer, Mathias & Detrano, Robert. *Heart Disease Dataset*. UCI. archive.ics.uci.edu/ml/datasets/heart+Disease

10. Rajalakshmi, R., Subashini, R., Anjana, R. M., & Mohan, V. (2018). Automated diabetic retinopathy detection in smartphone-based fundus photography using artificial intelligence. *Eye (London, England)*, *32*(6), 1138–1144. https://doi.org/10.1038/s41433-018-0064-9

11. Alaa, A. M., Bolton, T., Di Angelantonio, E., Rudd, J., & van der Schaar, M. (2019). Cardiovascular disease risk prediction using automated machine learning: A prospective study of 423,604 UK Biobank participants. *PloS one*, *14*(5), e0213653. https://doi.org/10.1371/journal.pone.0213653

12. Jahangiry, L., Farhangi, M. A., & Rezaei, F. (2017). Framingham risk score for estimation of 10-years of cardiovascular diseases risk in patients with metabolic syndrome. *Journal of health, population, and nutrition*, *36*(1), 36. https://doi.org/10.1186/s41043-017-0114-0

13. Kaggle Diabetic Retinopathy Detection competition. https://www.kaggle.com/c/diabetic-retinopathy-detection. Accessed November, 2018.

14. Gulshan V, Peng L, Coram M, et al. Development and Validation of a Deep Learning Algorithm for Detection of Diabetic Retinopathy in Retinal Fundus Photographs. *JAMA.* 2016;316(22):2402–2410. doi:10.1001/jama.2016.17216

15. Xie J, Ikram MK, Cotch MF, et al. Association of diabetic macular edema and proliferative diabetic retinopathy with cardiovascular disease: a systematic review and meta-analysis. *JAMA Ophthalmol*. 2017;135:586-593.

16. Dokken, B. (2008). The Pathophysiology of Cardiovascular Disease and Diabetes: Beyond Blood Pressure and Lipids. *Diabetes Spectrum.* Jul 2008, 21 (3) 160-165; DOI: 10.2337/diaspect.21.3.160

17. Voets, M., Møllersen, K., & Bongo, L. A. (2019). Reproduction study using public data of: Development and validation of a deep learning algorithm for detection of diabetic retinopathy in retinal fundus photographs. *PloS one*, *14*(6), e0217541. https://doi.org/10.1371/journal.pone.0217541